# Plasmonics of opalic surface: a combined near- and far- field approach

Clotilde Lethiec[1], Guillaume Binard[1], Trajan Popescu[2], Hugo Frederich[1], Phan Ngoc Hong[1,3], Eduardo Yraola[1,4], Catherine Schwob[1], Fabrice Charra[2], Laurent Coolen[1], Ludovic Douillard[2]*, Agnès Maître[1]*

[1]INSP, CNRS7588, Université Pierre et Marie Curie-Paris 6,
4 place Jussieu, F-75252 Paris cedex 05, France
²SPEC, CEA, CNRS, Université Paris-Saclay
CEA Saclay F-91191 Gif sur Yvette, France
[3]IMS, VAST, 18 Hoang Quoc Viet Road, CauGiay District, Ha Noi, Viet Nam
[4]Dpto. Física de Materiales, Universidad Autónoma de Madrid
Campus de Cantoblanco, 28049-Madrid, Spain

**Abstract**

An opalic plasmonic sample, constituted by a hexagonal arrangement of metallized silica spheres, presents remarkableoptical properties due to the mixing of periodic arrangement and singularities at the sphere touching points. It is therefore an interesting candidate for exploiting the excitation of both localized and propagating surface plasmons. Several channels of excitation based on these properties orexploitinga certain level of disorder are evidenced, openingnew routes for the efficient excitation of plasmons on a wide spectral range. The versatility of such hybrid system is evidenced in the context of two complementary experiments: specular reflective spectrometry and photoemission electron microscopy. Both techniques offer different points of view on the same physical phenomenon and the link between them is discussed. Such experiments evidence the opportunities offered by these 2D hybrid materials in the context of nanophotonics.

Engineering the density of states is a key issue in nanophotonics for controlling and manipulating the interaction between light and matter. Passive devices like waveguides as well as active ones like light nanosources require the manipulation of the density of states in volume or on surface.

Photonic crystals are well known for the engineering of the Q factor. In 2D structures high Q factors are obtained in connection to the use of high contrast index materials[1,2,3,4,5]. In3D,photonic crystalsare usually fabricated with low-index materials yielding tolower Q factors.However, laserinduced lithographic crystals[6,7] or self-assembled artificial opals[8,9,10] have permitted the engineering of the density of states[11],leading to a reinforcement of the light-matter interaction for emitters inserted in these structures[12,13,14,15].Despite the limitation in achievable Purcell factors in low-Q cavities, a modification of emission rate has been evidenced[16,17]. A major characteristic of low-Q nanostructures is that they are much less demanding in terms of spectral matching. They offer a high versatility in the spectral tuning of the devices which isa major advantage for applications. Another strategy for increasing the interaction between light and matter, is not only to play with the Q factor, but also with the spatial confinement of the mode. Therefore plasmonic devices offering the opportunity to obtain intense fields in a very small volume are very good candidates for tuning the emission. An acceleration of the spontaneous emission in various antenna devices has been evidenced[18,19,20,21]as well as an increase in the photon extraction[22,23,24].Moreover, the versatilityof plasmonic samples due to their large spectral response is reinforced for 2D materials, as they offer easier manipulation for coupling with other components, such as emitters or fluidic circuits, andexpands the range of available imaging and characterization tools.

Plasmonic 2D materials offer different channelsfor plasmon excitation:on one hand disorderedAufilmsclose to their percolation threshold have been demonstrated to be interesting systems for excitation of localized plasmons due to coherent multiple scattering in this random system[25]. They also take advantage of the singularities of localized gold structures at percolation together with the broadband properties induced by the disorder. On the other hand excitation of surface plasmons polaritons is ruled by the matching between their wave vectors and the one of incident light. This can be achieved either with gratings[26, 27] or with slits or plasmon launchers[28].

In this paper, we propose the use of a plasmonic opal surface (i.e. an opal, or fcc structure of dielectric spheres, covered with gold) as a 2D versatile plasmonic surface. Such a plasmonic sample is a combination of a 2D grating (a compact plane equivalent to the(111) plane of afcc crystal) and a set ofgeometrical singularities at the microscopic scale at the contact pointsbetween adjacent spheres. This geometry allows competition betweenlocalized plasmonic modes (LSP) and propagating surface plasmon polariton modes (SPP) in a broad spectral range. Moreover the level of disorder in this periodic structure is easily tunable by controlling the size distribution of the spheres during the synthesis process. Thus it opens new opportunities of light matter coupling. In order to gain a wide knowledge of the plasmonic properties of such a material, two different analysesareexploited. Specular reflectivity spectrometry evidencesspatially integrated absorption due to plasmon excitation whereas photoemission electron microscopy images the excited plasmons at the microscopic scale.

**RESULTS AND DISCUSSION**

**Sample preparation- Artificial metallic opals**

Our objective is to design a specific sample allowing for a broadband and efficient excitation of both localized and propagating plasmons on different channels. Metallized self-assembled colloidal crystals, allowing for the excitation of grating induced plasmons, the creation of plasmon launchers, and a controlled level of disorder are excellent candidates for these purposes.

Artificial metallic opals are prepared by evaporatinga Au layer on a self-assembled periodical array of silica spheres (opal) obtained by the convection method[29, 30]. The technique[31]consists in the heating of a solution of beads diluted in ethanol at 25°C, in which a glass coverslip is immersed in a vessel. Upon solvent evaporation, the beads self-organize on the glass substrate at the meniscus position, leading to a centimeter sized opal sample. The opal lattice is face-centred cubicfcc, with its (111) plane parallelto the sample surface.In a second step, a 150 nm thick silica buffer is deposited on top of the opal surface by electron beam evaporation so that the opal corrugation is smoothed. A Au film is then evaporated on this latter surface. The Au layer thickness of about 250 nm is larger than the SPP skin depth, so that the metal layer is optically thick. Figure 1 presents a schematic 3D view of the artificial Au-covered opal used in this work.

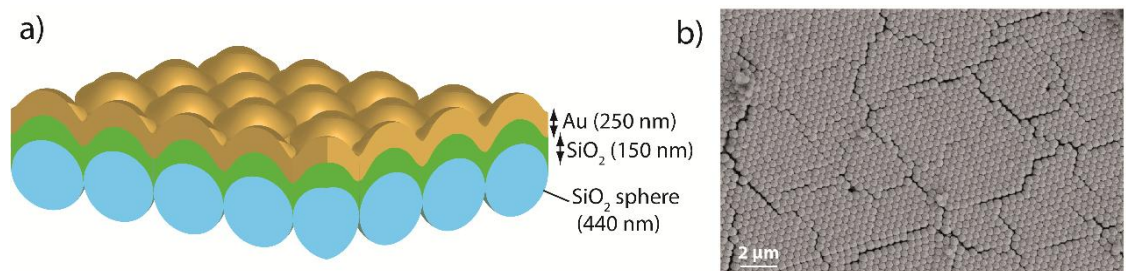

**Figure 1.** a) Schematic 3D view of the artificial Au-covered opal. SiO$_2$sphere diameter 440nm, SiO$_2$ smoother layer 150 nm, on top Au layer 250 nm.b)SEM image of the Au-covered opal, size = 12μm x 8μm

A *scanning electron microscope* (SEM) image of the opal surface is presented in figure 1.b. The opal surface is predominantly covered by a conformal Au layer, as observed in the case of Ag deposition[32]. The lattice is hexagonal compact in agreement with afcc (111) terminating plane. A network of cracks, related to the opal drying conditions, is also clearly identifiable. These cracks, about 815± 130 nm in width, determine grains of typical size 6.4± 1.8 µm in diameter. The crystallographic orientation of the surface is maintained from one grain to the other.

Figure 2.a presents an *atomic force microscope* (AFM) image of a crack free region of the metallic opal surface. The measure of the apparent sphere diameter gives a = 440 ± 9 nm. The average height between the highest peak and lowest valley as extracted from AFM profile corresponding to the average groove depth between two adjacent beads (figure 2.b) is 2h = 83± 3nm. The latter parameter defines the groove depth of the opal profile.

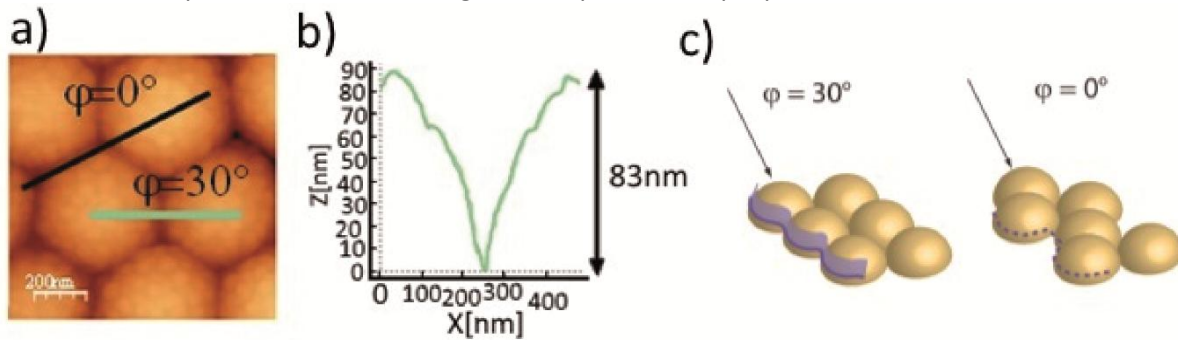

**Figure 2.** a) AFM image of an Au-covered opal, on a crack free grain region. φ is the azimuthal orientation b)Surface profile corresponding to the straight green line on the AFMtopography image. c)views of the surface along the 2 different direction φ=0° and φ=30°. A line is drawn along the spheres at half the groove depth. The surface tangent to the sphere and containing this line is in the case of φ=30° close to a plane tilted with respect to the horizontal plane. On the contrary, for φ=30°, the line is sinuating with a large amplitude, so that the tangential surface cannot be assimilated to a plane

**Plasmonic modes –Resonant diffractive excitation of SPPs**

In a first set of experiments, plasmons are optically excited and the coupling between incident light and plasmons is analysed by optical polarization resolved reflectometry[23, 31]. A polarized quasi-collimated (around 1-mm of diameter) low irradiance power halogen white-light beam is directed onto the sample with an incident angle θ with respect to the surface normal, collected by an optical fiber at specular angle and analyzed by a spectrometer.All spectra are normalized by the spectrum of the illuminating lamp.The azimuthal angle φ corresponds to the angle between the plane of incidence and the opal main crystallographic axis (φ = 0° when $k_{//}$, the projection of the wave vector on the surface, is along the direction [11]). .

Figure3displays the reflectivity spectra measured on anAu-covered opal. For excitation light along the azimuthal angles φ = 0°, spectra in p and s polarisation exhibit different evolutions with the change of incident angle. We first discuss the results for p-polarized incident light.

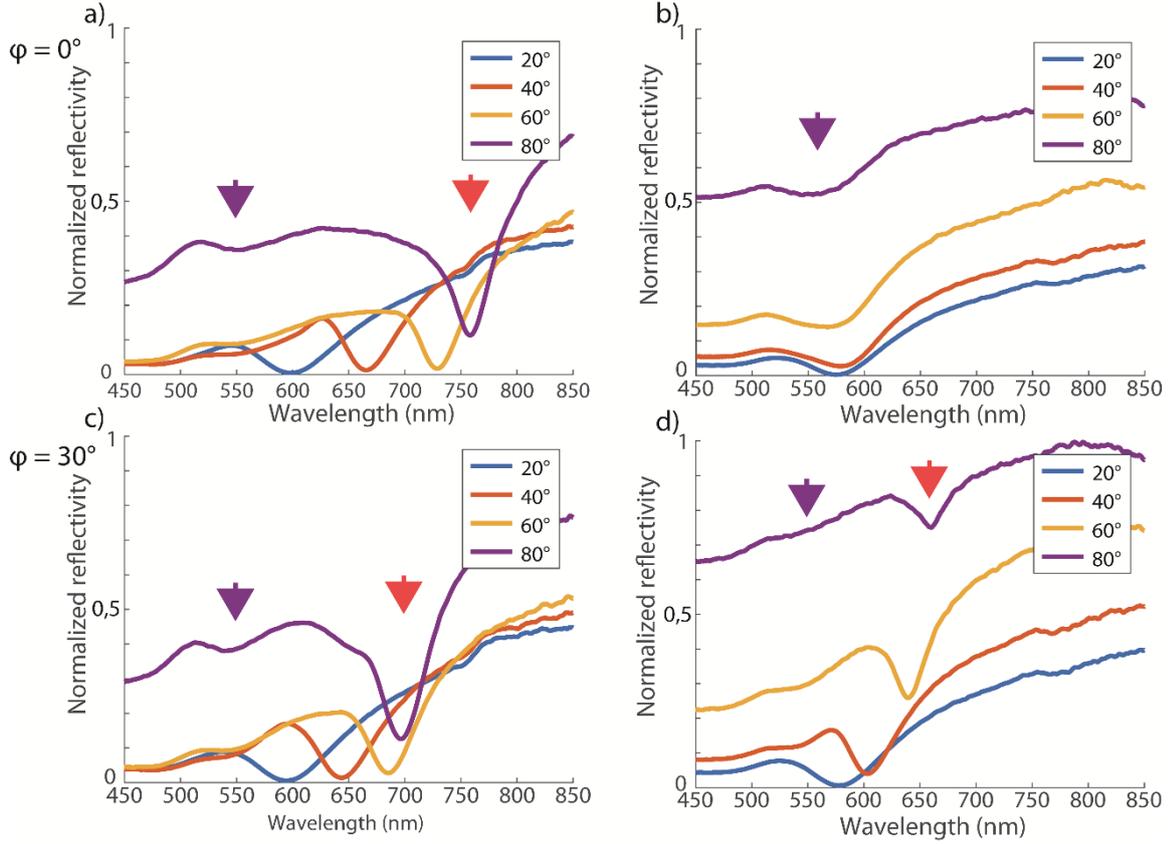

**Figure 3.** Normalized reflectivity spectrain p (left) and s (right) polarization respectively, as a function of the excitation wavelength, for an incident angle θ between 20° and 80°, and for two azimuthal angles φ = 0° and φ = 30°

In p polarization (left panel), a large dispersion of the main absorption dipwith incident angle is observed between 600 nm and 750 nm: the reflectivity dip moves toward longer wavelengths as the incident angle increases. The reflectivity minimum value amounts to only a few percents of the maximum one. The p-polarized reflectivity curves are qualitatively close for different azimuthal angles φ. These features constitute a direct signature of the efficient and resonant diffractive coupling of light with transverse magnetic SPP waves propagating at the opal surface[31].

In order to achieve efficient diffractive coupling between the incident propagating wave (wave vector $\vec{k}_0$ and incident angle θ) and propagating surface plasmon polaritons, the following phase matching condition, referred to as resonance in the following, has to be fulfilled:

$$\vec{k_{SPP}} = \vec{k}^0_{//} + \vec{G_i} \qquad (1)$$

With $\vec{k}^0_{//} = \vec{k}_0 \sin(\theta)$. $\vec{G_i}$ are the vectors of the reciprocal space of the fcc (111) opal surface, of which norm G scales as $G = \frac{4\pi}{\sqrt{3}a}$. This phase matching condition just expresses that the mismatch between plasmons and incident propagating waves can be compensated by the reciprocal grating vectors. By introducing the azimuthal angle φ, this dispersion relation scales as:

$$k_{SPP}^{h=0} = k_0 \sqrt{\frac{\varepsilon_d \varepsilon_m}{\varepsilon_d + \varepsilon_m}} = \left| k^0_{//} \cos\varphi - m \frac{4\pi}{\sqrt{3}a} \right| \quad \text{with } \varphi = [0, \pi], \text{ m an integer} \qquad (2)$$

This condition is fully valid for a quasi-flat metal / dielectric interface, i.e. at the zeroth order in groove depth h[31]. For any significant corrugation, further physical phenomena have to be considered. When the azimuthal angle φ increases up to 30°, SPP excitation is blue shifted as shown in figure 3.c, in qualitative accordance with the evolution of the theoretical phase matched relationship (2).

Due to the hexagonal geometry of the surface and the rotational periodicity of 60°, we restrict our study the variation of φ to the range [0°, 30°].

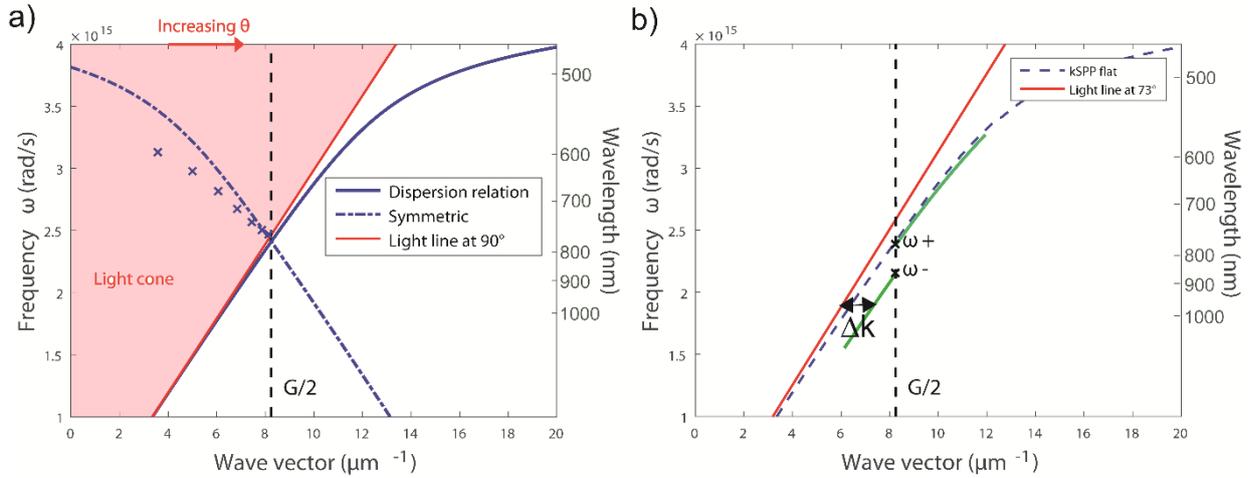

**Figure 4.a)** Calculated dispersion relation of SPP modes as a function of $k_{//}$ on a flat Au surface (solid blue line), and its symmetric (dashed blue line) representing the folding in the first Brillouin zone for a metallic grating of periodic vector G. The light cone (in pink) limited by the light line (in red) represents respectively all propagating modes in free space and the mode incident on the surface at θ=90°. The blue points correspond to the experimental dips of p polarization specular reflectivity. Points obtained at large incidences are the closest to the light line. **b)** light line at θ=72.5° (red line), dispersion relation of SPP modes on a flat Au surface (blue dashed line), ω+ and ω– limits of the gap opening for a low corrugated sinusoidal surface, parallels to the dispersion relation through points (G/2, ω+/–) (green lines)

In Figure **4.a,** the dispersion relations of SPPs at a flat Au/vacuum interface are plotted on both sides of the k=G/2 limit. On the same figure, the experimental points corresponding to the maxima of absorption for a p polarization and φ = 0° are superimposed. These points are found to be in qualitatively good agreement with the SPP dispersion relation, especially for θ close to 90°. Close to the normal and the grazing incidences, the opening of a gap should be accounted for in the model.

So, by acting on the two independent angular parameters φ and θ of the excitation orientation, the opalic plasmonic surface, allows for an efficient and tunable resonant diffractive coupling to SPP modes.

**Opening of a gap**

On a corrugated surface a frequency gap opens at the first Brillouin zone boundary for $k_{SPP} = \frac{G}{2} = \frac{2\pi}{\sqrt{3}a}$ where two symmetrical plasmon branches form an anticrossing. The corresponding frequency ω will be noted: $\omega = \omega_0$. For weak sinusoidal surface profile (corrugation smaller than 80nm), which approximates the first level of our cycloid surface, the gap opening on the SPP dispersion curve can be described analytically[33, 34, 35] within a perturbative approach. Two points need to be mentioned, (i) first the normalized gap width increases with the corrugation amplitude (for corrugation smaller than 80nm), (ii) the central position of the gap $\bar{\omega}$ decreases as the corrugation amplitude rises.

$$\left(\frac{\omega_+}{c}\right)^2 - \left(\frac{\omega_-}{c}\right)^2 = \frac{4G^2}{\sqrt{-\varepsilon_m \varepsilon_d}}(Gh)\left(1 - \frac{7}{2}(Gh)^2\right) \qquad (3)$$

$$\frac{1}{2}\left[\left(\frac{\omega_+}{c}\right)^2 + \left(\frac{\omega_-}{c}\right)^2\right] = \left(\frac{\omega_0}{c}\right)^2\left(1-(Gh)^2\right) \qquad (4)$$

For the experimental parameters a = 2R = 440 nm, 2h = 83 nm, expressions (3) and (4) yield a gap width of 84 nm (151meV) and a fall of the central position of $\bar{\omega}/\omega_0$ = 0.940 with ($\bar{\omega}$, $\omega_0$) = (2.27 $10^{15}$ Hz, 2.42 $10^{15}$ Hz) = (830, 779 nm). The predicted top and lower bandgap edge values are ($\omega_+$, $\omega_-$)=(2.38 $10^{15}$ Hz, 2.15 $10^{15}$ Hz)corresponding in terms of wavelength to ($\lambda_+$, $\lambda_-$)=(790, 874 nm). For the calculations, the Au dielectric constant as determined by Johnson& Christy[36]is considered. The positions of the lower and higher frequencies limiting the gap are reported in figure **4.**b). This gap can be observed in figure 4a at high incidence (crossing of the $k_{//}$ = G/2 line) and at low incidence ($k_{//}$ = 0 for larger frequencies) where experimental data deviate (by a few percent) from the theoretical phase matching model. We can notice as well in figure 3 that resonant excitation above 760 nm could not be obtained in accordance with the gap value calculated here.

**Resonant coupling to s-polarised modes and to localized plasmon modes**

The reflectivity spectra in s polarization are plotted in the right panel of figure 3.For $\varphi = 0°$, reflectivity curves present a completely different feature as compared to p polarization. They exhibit one modest slightly dispersive low reflectivity dip, at 570nm, far less pronounced than the one observed with p-polarized light. This dip is assigned to localized surface plasmon LSP modes associated to geometric gaps at touching sphere locations[31].

It can be noted that when the azimuthal angle $\varphi$ is increased a second dispersive mode appears for s-polarization with a maximum of dispersion obtained at $\varphi = 30°$. As shown in figure 2c) along this direction the incident light impinges not only onto the horizontal surface but as well onto a smoothly corrugated surface (drawn in purple) which can be assimilated to a plane tilted with respect to the horizontal. Indeed the incident light s-polarized with respect to the horizontal plane acquires a p-component with respect to the tilted plane. This favours a diffracting coupling to the SPP modes and can explain the dispersive mode excitation observed for s-polarized incident light propagating along the $\varphi = 30°$ direction. On the contrary, as described in figure 2c), this argument is not valid along the $\varphi= 0°$, for which no tilted plane can be defined: only LSP excitation could be observed.

Localized plasmons could also be observed in p polarization. At grazing incidence, when SPP excitation has switched to NIR, a small dip around 560 nm appears (violet arrow), corresponding to the excitation of LSP modes between adjacent balls, just as described for s polarization. This latter conclusion will be substantiated in the next section.

**Plasmonic modes – Off resonance excitation of SPPs at surface discontinuities**

In a second set of experiments another point of view will be adopted. The sample is illuminated in the near infrared (NIR) by a Ti:Al$_2$O$_3$ femtosecond laser and analyzed by photoemission electron microscopy (PEEM).Instead of considering overall light absorption due to plasmon excitation like in former reflectometry experiment, PEEM image the electromagnetic local density of states of opalic plasmons whatever their excitation origins, i.e. by on or off resonance excitation channels[37, 38, 39, 40].This mapping of the coupling between light and surface plasmon gives routinely access to full-field images of the surface optical near field distribution, at a scale which canbe much smaller than the illuminating laser spot (700 μm) down to below the sphere diameter[33, 34, 35].This near-field resolution offers additional information to the spatially integrated far-field reflectivity measurements.

Unlike the former measures, photoemission experiments are performed at a fixed incidence angle, θ = 72.5° ± 2.5° with respect to the surface normal, while the photon wavelength is continuously tuned.

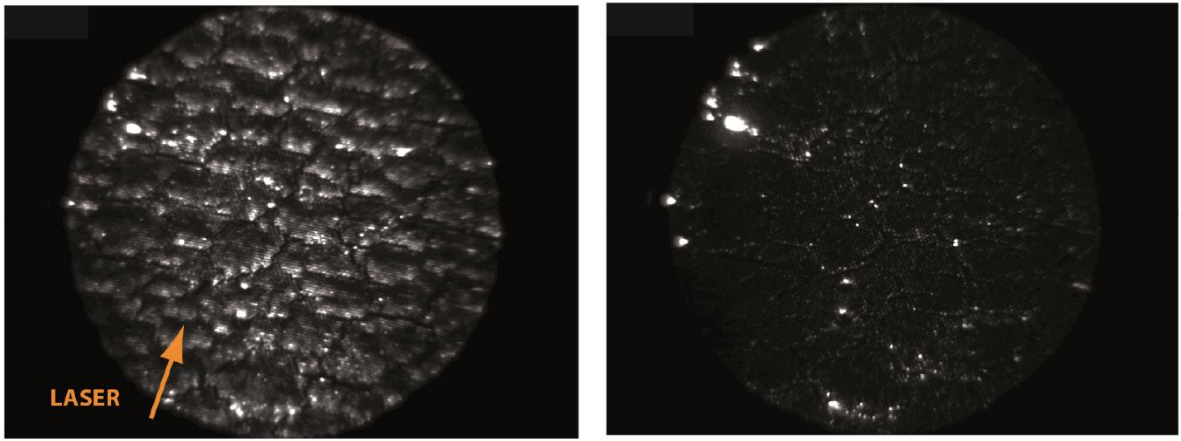

**Figure 5.** PEEM images of the same area of the opal sample under 800 nm illumination in p (left) and (s) (right) polarization respectively. The direction of the laser beam corresponds to the marked arrow ($\varphi = 0°$) (same on both images). Fields of view: 57 μm.

Figure 5 gives the general overview of the features of a cracked opal surface under light illumination ($\varphi = 0°$) as observed in large PEEM fields of view. In p – polarization, one observes domains about 6.4 μm in diameter separated by narrow cracks imaged as dark regions. At the scale of an individual domain, the photoemission signal exhibits a hexagonal lattice structure corresponding to the periodicity of the metallic grating. More noticeably, bright fringes can be observed perpendicularly to the beam direction for wavelengths in the range [800,910 nm]. From place to place, intense hot spots are visible either on the cracks or distributed over the surface sample. For wavelength above 910 nm and below 780nm, no fringes could be observed. In s - polarization, for any wavelength, PEEM images present just hot spots within the close packed arrangement of metallic spheres.

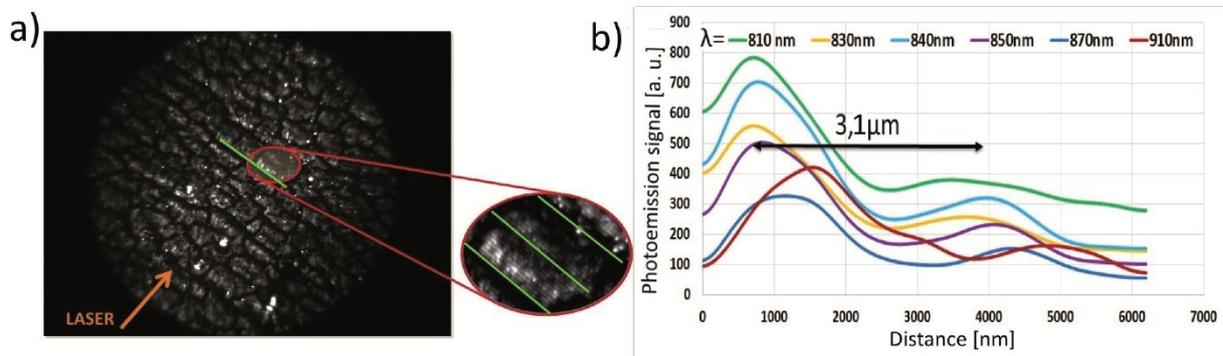

**Figure 6.** a) PEEM Image of a 440nm opal under 840 nm illumination in p-polarization The electric field is higher at the leading edge of each individual grain right after the crack front b) Line profile at different distances ofone selected domain along the beam direction for wavelengths ranging from 800 nm to 910 nm. Further details can be found in supporting information

Upon p-polarized light illumination in the wavelength window [800, 910 nm], a well-defined stationary-wave pattern can be observed on the opalic surface. As visible in figure 6.a, interferences appear from the leading edge of each individual grain perpendicularly to the incident laser beam direction (orange arrow). The period of the fringes (figure 6.b) keeps a constant value of 3.1 ± 0.2 μm over the wavelength window and their maxima move toward the core of the grain as the wavelength increases. Their intensities are maximal for an excitation wavelength around 810 nm. The overall intensity profile decreases toward the centre of the grain with an attenuation length measured as 11.3 ±1.6 μm.

This interference pattern constitutes an experimental evidence of the excitation of propagating SPPs which interfere with the full field illuminating light giving rise to a beating pattern[41, 42, 43]. Indeed, in this range of wavelength, as shown in figure 4.a and b, no phase matching can be directly achieved between the incident laser under $\theta = 72.5°$ and the SPP neither for the flat interface branch nor for the folded one. Therefore, unlike demonstrated in specular reflectometry experiment, these oscillating features cannot be explained by grating induced resonant excitations of SPPs. Their origin can be attributed to the existence of the crack network (mesoscopic disorder) which opens a new channel of light matter coupling in addition to the resonant diffractive process previously described. Indeed, cracks constitute sharp discontinuities within the opal close packed structure which break the translational invariance and thus play the role of SPP launcher sites. At grain boundaries, the missing momentum between light and SPP waves can be brought by the crack profile and fulfilling of the phase matching condition gives birth to SPP waves. For a crack-free opal surface, we observed no interference phenomenon at any wavelength and the photoemission signature is limited to a large collection of hot spots associated to LSP modes taking place at the geometrical gaps between touching spheres[44]. Such mesoscopic disorder and multiple drying cracks on the opalic surface arethen responsible for the opening of a new type of SPP excitation channel beyond the diffractive one.

The beating period can be deduced from the dispersion diagram of figure 4b byevaluating the differences of momenta between light and SPP waves $\Delta k = k_{SPP} - k_{//} = 2\pi/\lambda_{beat}$. Considering analytic dispersion of SPP on a flat Au/vacuum interface, one obtains a period of 13,8 µm at $\lambda_0$ = 874 nm which exceeds significantly the 3.1 µm value measured experimentally.However, as mentioned previously, the periodic corrugation of the surface induces in the SPP dispersion relation, at the boundary for $k_{SPP}$ = G/2, the opening of a plasmonic frequency gap. In figure 4.b, we represent not only the light line and the SPP dispersion curve but also the ($\omega_-$, $\omega_+$) frequencies corresponding to the opening of the gap and calculated by the low corrugation sinusoidal model previously introduced[33, 35]. As a rough approximation, in the vicinity of ($\omega_-$, $\omega_+$), we interpolate the dispersion branches as a line parallel to the SPP dispersion curve (see methods section). It has to be noticed that far from the gap edges, the dispersion relation is no more affected by the gap and can be approximated by the SPP dispersion relation on the flat Au/vacuum interface, which will lead to a larger value of the beating period.

For the lower frequency branch $\omega_-$ probed by the PEEM experiment, atan excitation wavelength of 874 nm (predicted lower bandgap edge wavelength), the increase of the momenta difference due to the opening of the gap leads to a decrease of the analytic beating period down to $\lambda_{beat}$ = 5.07 µm in better semi-quantitative agreement with the experiment. In the limits of this model, this beating period remains quite constant in the lower branch $\omega_-$ ($\omega < \omega_-$ and $\lambda > \lambda_-$). Experimentally, we observe the same evolution of the beating period which remains close to 3.1 µm in the whole range of observation of the fringes [800, 910nm]. By following this argument, we could estimate that the bandgap appears for wavelengths lower than 800nm, at frequencies slightly higher compared to those predicted by the model.

Hence, the gap opening model[33] developed here shows a good qualitative agreement with our experimental results, but encounters some limitations in the exact quantitative description. The latters can be attributed to several factors. The sample exhibits a 2D corrugation instead of the 1D profile used in the modeling. Then, the opalic surface profile is closer to a cycloid than to a smooth sine function. Finally, corrugation amplitude of our geometries sets at the limit of validity for the model. Nevertheless, this model sheds light on some of the features encountered for SPP excitation on plasmonic opals and provides a rather good quantitative agreement.

For wavelengths longer than 910nm, only hot spots and no fringes could be observed. This spectral range is no more in the vicinity of the gap edge and the density of states is reduced. Indeed, in the vicinity of the gap (flat band structure), the high density of states, permits not only the excitation of SPP but also to get a large PEEM response. Therefore, further in the IR domainwhere the density of states is lower, the fringes are not observed.

For wavelength shorter than 790nm, fringes are not observed either. This is probably due to the combination of several effects. In the gap region, propagation of SPP is not possible, preventing the observation of fringes. On the high energy side of the gap, at its vicinity, even though the local density of states is large, the expected beating period is large. The smooth sinusoidal model considered here leads to a value of $\omega_+ = 2{,}38\ 10^{15}$ Hz corresponding to $\lambda_+ = 790$nm, the corresponding beating period amounts to $\lambda_{beat} = 9.5$ µm. By considering a 75 nm blue shift as suggested by experiment for the lower bandedge value, one gets a top bandedge wavelength of $\lambda = 715$nm, and a beating period on the order of $\lambda_{beat} = 7.3$ µm wavelength. These values both exceed the 6.4 µm typical lateral dimensions of the cracked opal domains, preventing any observation of the beating figures. Finally far from the gap, at high energy in the visible, where the expected beating period gets smaller than the opal domain size, SPP excitation and propagation suffer both from a lower density of states and a high absorption by the gold medium, which limits the propagation length and any experimental detection.

The profile of the interference pattern gives also access to the SPP propagation length at the corrugated opal surface. Experimentally, at 856 nm, we obtain $L_{Prop.}$(opal, 856 nm) = 11.3 +/- 1.6 µm. This measure is small compared to that calculated for a flat Au/vacuum interface $L_{Prop}$(Au/vacuum, 855 nm) = 60.0 µm[45]. This short length can be explained by the surface corrugation. Indeed, theoretical calculations for SPP waves propagating along a hexagonal arrangement of (non touching) Au hemiellipsoids (a = 450 nm) defines a superior limit of the propagation length of 36 µm in better agreement with the experimental data[46].

**Plasmonic modes – Excitation of LSPs at surface gaps**

For p and s polarizations, hot spots are obtained over the whole spectral range, with the specificity that for s polarisation, no propagating plasmons wavescould be efficiently excited. In the NIR range, hot spots are obtained all over the spectral range with themost intense ones corresponding to structural defects like cracks on the surface. The number of hot spots decreases when the wavelength increases, a behaviour reminiscent ofthat of a random metallic film[25].

In the visible range, hot spots are observed for both polarizations in agreement with previous reflectivity measurements. It is now confirmed through PEEM experiment, that it also corresponds to the signature of LSP excitation at the geometrical gap between adjacent spheres. This will be developed in a forthcoming publication.

**CONCLUSIONS**

In summary, we compared and discussed the connections between specular reflectivity spectrometry and photoemission electron imaging. While the first one favours resonant excitation of SPPs and relies on a spatially integrated light absorption, the electron microscopy images, at a subwavelength resolution, the electromagnetic field distribution of any plasmonic excitation. In other words, reflectivity spectrometry relies on the evidence of the origin of the plasmons excitation (absorption of incident light), whereas photoemission images, at small scale, the electronic density of state which is the consequence of plasmon excitation. Reflectivity spectrometry offers a large versatility permitting full exploration of dispersion lines by simple light – sample geometry adjustment at a cost of a spatially integrating technique. Photoemission electron microscopy is a more demanding technique mapping both on and off resonance plasmonicmodes at a cost of a local description. Hence, both methods are complementary as they offer different points of view.

Broadband excitation of plasmons permitting the competition between propagating and localised plasmons has been evidenced. It relies on excitation of plasmons by different channels. Resonant diffractive excitation of SPPs is achieved by reflectivity measurement, whereas PEEM permits the additionalinvestigation of off resonanceSPP excitation thanks to the high pumping level and the presence of cracks acting as plasmon launcher sites. SPP propagating wave in a periodic medium have been visualized and evidenced as fringes from photoemission microscopy data and as dispersion lines and bandgap from both methods. The excitation of propagating plasmons on the surface is associated to a high density of electromagnetic states highly dependent on the resonance with the dispersion line and on the presence of the gap.

Hot spot excitation has been demonstrated in all the visible – NIR range, and corresponds either to the excitation of localized plasmons at the location of geometric gaps between adjacent spheres, or to extremely intense hot spots on structural defects. So, plasmonic opals possess not only the advantage of presenting periodicity but also a certain amount of disorder, which permits a large tunabilityin plasmon excitation modes, either propagating as well as localized ones. Along this line we evidence several channels of excitation at the origin of a wide excitability over a large spectral range at different operating conditions.

Combination of emitters with low efficiency and this opalic sample are promising devices. Thanks to the high electromagnetic density of states on the surface and especially at hot spots locations, emission could be enhanced by a large factor. Moreover, propagating plasmons can permit the transfer of information from one spot to the other. In a forthcoming paper, the interplay between SPPs and LSPs will be evidenced and discussed[44]. The features investigated in this paper like the combination of order (periodic grating) / disorder (mesoscopic crack network), the presence of numerous channels of plasmon excitation (diffractive coupling, plasmon launcher site, on and off resonance, structural defects, geometrical gaps between adjacent balls), the large spectral range of dispersive and non-dispersive excitation, make plasmonic opalic sample a versatile 2D platform for plasmonics and nanophotonics.

**METHODS**

**Imaging techniques.** The SEM imaging was performed on a ZEISS Supra 40*scanning electron microscope.* The *atomic force microscopy*images were obtained on a JPK AFM instrument.

**Optical specular reflectometry**A collimated white-light beam passed through a 1-mm diaphragm and a polarizer and is directed onto the sample with the incident angle θ. The reflected beam is collected at the same angle θby an optical fiberthrough a 2-mm diaphragm (chosen larger than the former diaphragm because of the slight beam divergence) in order to separate reflection from low-angle scattering. An angular resolution of 2° is achieved.For the normalization of the spectra, the spectrum of the illumination lamp is measured with the 2-mm diaphragm and the optical fiberfacing the incident beam. The range of the incidence angle θ is [20°, 80°]. The excitation is made either by a halogen lamp (spot width on the sample: 1.5 mm) or by a supercontinuum laser (spot width: 700 µm). The excitation polarization is set in p or s polarization state by placing a Polaroid film in the laser path.

**Photoemission microscopy investigation.** Mapping of the near optical field at subwavelength resolution is obtained by photoemission electron microscopy. The sample is illuminated by a femtosecond pulsed laser (Ti:sapphire ultrafast oscillator Chameleon Ultra II, Coherent Inc., repetition rate 80 MHz, hyperbolic-secantsquared pulse, pulse width 140 ± 20 fs). The polarization of the laser light is controlled by a half-waveplate. The fixed incidence angle is θ = 72.5° ± 2.5° with respect to the surface normal. The accessible wavelengths cover two regions (i) the [690 nm, 1040 nm] near-infrared range and (ii) the [520 nm, 640 nm] visible range through the use of an *optical parametric oscillator* OPO (APE). The PEEM instrument used is a commercial model (Elmitec LEEM/PEEM III) which operates under ultrahigh vacuum conditions. On a flat surface lateral spatial resolutions of 10 nm and 20 nm are routinely achieved in LEEM (*low energy electron emission*) and PEEM imaging modes, respectively. The imaging principle makes use of the exalted electron emission at near field hot spots through a multiphoton light absorption process. For Au, an apparent 3-photon photoemission regime is measured in the wavelength window [550 nm, 1040 nm]. The experimental conditions are optimized to keep constant the intensity, the pulse width and the focus position of the incident beam during data acquisition. The latter is carried out at a low sample temperature of about 125 K to avoid any sample damage upon laser light illumination.

In PEEM imaging mode, the measured brightness in a given image area is proportional to the electron emission from this area. For quantitative data extraction integrated PEEM signal is computed in two steps: i) CCD background subtraction, ii) integration of the pixel values over a representative image area.

**Fringe spacing.** Figure 6 shows the line profile of the photoemission signal along the beam direction within one crack free domain for wavelengths in the range [800, 910 nm]. For each wavelength, the line profile data are obtained in a two-step image processing sequence. In a first step, the different accessible line profiles are averaged along the direction of the fringes to extract an average photoemission profile specific of the whole selected domain. In a second step, these data are further smoothed according to a 5 points Savitzky-Golay filter to remove the high frequency signal associated with the individual sphere positions acting as discretization acquisition points.

In figure 4b, the dispersion curve (green line) in the lower (resp upper branches), is evaluated by considering a linear dispersion starting with (G/2, $\omega_+$) point (resp ((G/2, $\omega_-$)) . This linear evolution is a rough approximation of the real dispersion curve not strictly valid at the gap and far from it, but in the vicinity of the gap, it gives good order of magnitude of the beating period between incident light and SPP modes.

Figure S1 displays the fringe period $\lambda_{beat}$ vs the illuminating wavelength. For each wavelength, the beating period is determined from the dispersion diagram of the corrugated opalic surface (figure 4b) by evaluating the differences of momenta between light and SPP waves $\Delta k = k_{SPP} - k_{//} = 2\pi / \lambda_{beat}$ by considering either a flat interface (dashed line) and both the upper and lower branches of the corrugated surface (solid lines).

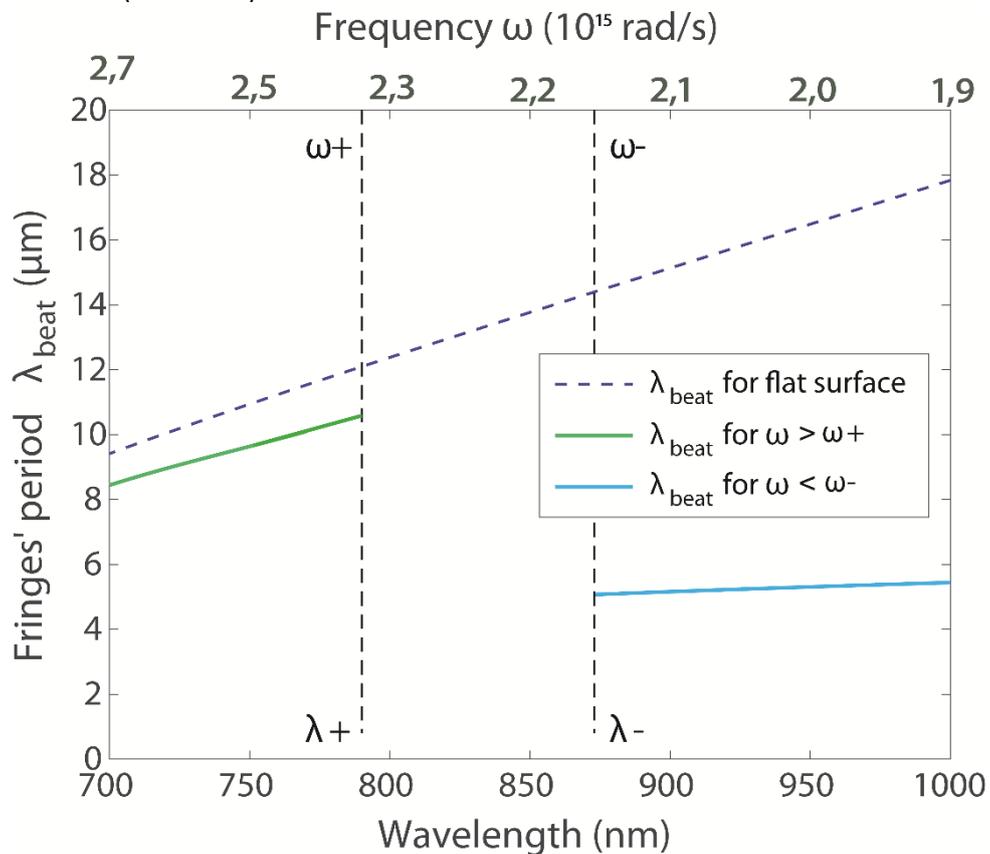

**Figure S1:** Fringe period $\lambda_{beat}$ versus the illuminating wavelength. The dashed line corresponds to the fringe period calculated by considering the beating between the incident light and SPPs excited at a flat surface (dashed line). The green (rep blue) solid line corresponds to the fringe period from the upper (resp lower) branch as explained above.

Here are given all the values of the parameters of the gap model for a sphere diameter of 440nm and a groove depth of 83nm:

| | |
|---|---|
| $\omega_0 / \lambda_0$ | $2,42.10^{15}$ rad.s$^{-1}$ / 779 nm |
| $\bar{\omega} / \bar{\lambda}$ | $2,27.10^{15}$ rad.s$^{-1}$ / 830nm |
| $\Delta\omega / \Delta\lambda$ | $2,29.10^{14}$ rad.s$^{-1}$ / 84 nm |
| $\omega_+ / \lambda_+$ | $2,38.10^{15}$ rad.s$^{-1}$ / 790 nm |
| $\omega_- / \lambda_-$ | $2,15.10^{15}$ rad.s$^{-1}$ / 874 nm |

**ACKNOWLEDGMENT.**
The authors want to thank Luisa Bausa, Willy Daney de Marcillac, Stephane Chénot, Loic Becerra, Dominique Demaille, FrédéricMerlet, Dominique Martinotti and Emmanuelle Lacaze for their technical and scientific supports.This work was supported by French state funds managed by the ANR within the Investissementsd'Avenir programme under reference ANR-11-IDEX-0004-02, and more specifically within the framework of the Cluster of Excellence MATISSE and by the Direction Générale des Armées (ANR-DGA "Calypso"). The CEA authors acknowledge financial support by the French National Agency (ANR) in the frame of its program in Nanosciences and Nanotechnologies (PEEMPlasmon project no. ANR-08-NANO-034), Nanosciences Île-de-France (PEEMPlasmonicsproject) and the "Triangle de la Physique" (PEPS project no. 2012 035T).